\documentstyle[qqaalart]{article}

\author{Jos\'e Antonio Belinch\'on\thanks{E-mail: abelinchon@caminos.recol.es}\\
Grupo Interuniversitario de An\'alisis Dimensional\\
Dept. F\'isica ETS Arquitectura UPM Juan de Herrera 4 28040 Espa\~na}
\title{Some FRW models with variable $G$ and $\Lambda $
}
\input tcilatex

\begin{document}

\maketitle
\begin{abstract}
We consider several models with metrics type FRW, with $k=0$ and with a
generic equation of state, $p=\omega \rho ,$ furthermore, we take into
account the conservation principle of the energy-momentum tensor, $%
div(T_{ij})=0,$ but with $G$ and $\Lambda $ variable. We find trivially, a
set of solutions through dimensional analysis. We stand out that, $G\propto
t^{1+3\omega },i.e.$ its behaviour depends on the state equation and $%
\Lambda \propto t^{-2}$ in any case. The scale factor $f$ varies as $%
f\propto t$ in the studied models i.e. have no horizon problem.
\end{abstract}

\section{\bf \ Introduction.}

Recently have been studied several models with metrics FRW in those which
are considered the ''constants'' $G$ and $\Lambda $ as dependent functions
of the time $t$ (\cite{A}). In this paper we want to emphasize as the use of
the Dimensional Analysis (D.A.) permits us to find \underline{in a trivial
way} a set of solutions to this type of models, but with $k=0,$ and taking
into account the conservation principle. To stand out also, that the
variation of $G$ that is obtained in this type of solutions is always
proportional to the time (depending on the state equation) and that $\Lambda 
$ always show a behaviour inversely proportional to $t^2$ i.e. $\Lambda
\propto t^{-2}$ independently of the state equation that we impose ($%
p=\omega \rho $ / $\omega =const.)$.

The paper is organized as follows: In the second paragraph we present the
equations of the model and are made some small considerations on the
followed dimensional method (address to reader to the classic literature on
the topic (\cite{B})). In the third paragraph we make use of the dimensional
analysis (Pi theorem) to obtain a solution to the principal quantities that
appear in the model and finally in the fourth paragraph we end with a short
recapitulation and succinct conclusions.

\section{{\bf \ The model}.}

The modified field equations are: 
\begin{equation}
\label{e0}R_{ij}-\frac 12g_{ij}R-\Lambda (t)g_{ij}=\frac{8\pi G(t)}{c^4}%
T_{ij} 
\end{equation}
and we impose that%
$$
div(T_{ij})=0 
$$
where $\Lambda $$(t)$ represents the cosmological constant. The basic
ingredients of the model are:

\begin{enumerate}
\item  The line element is defined by:%
$$
ds^2=-c^2dt^2+f^2(t)\left[ \frac{dr^2}{1-kr^2}+r^2\left( d\theta ^2+\sin
{}^2\theta d\phi ^2\right) \right] 
$$
here only we will consider the case $k=0$

\item  The energy-momentum tensor is defined by:%
$$
T_{ij}=(\rho +p)u_iu_j-pg_{ij}\qquad \qquad p=\omega \rho 
$$
\end{enumerate}

The field equations, already developed, in function of the metric $g\in
T_2^0(M)$ of type FRW are: 
\begin{equation}
\label{e1}2\frac{f\,^{\prime \prime }}{f\,}+\frac{(f\,^{\prime })^2}{f\,^2}%
=- \frac{8\pi G(t)}{c^2}p+c^2\Lambda (t)\ \ 
\end{equation}

\begin{equation}
\label{e2}\frac{(f\,^{\prime })^2}{f\,^2}=\frac{8\pi G(t)}{3\,c^2}\rho
+c^2\Lambda (t)\qquad \quad \ 
\end{equation}

\begin{equation}
\label{e3}div(T_{ij})=0\text{ }\Leftrightarrow \rho ^{\prime }+3(\omega
+1)\rho \frac{f^{\prime }}f=0 
\end{equation}
integrating equation (\ref{e3}), we obtain the following equation. 
\begin{equation}
\label{e4}\rho =A_\omega f^{-3(\omega +1)} 
\end{equation}
where $f$ represents the scale factor that appears in the metrics and $%
A_\omega $ is the constant of integration that depends on the state equation
that is imposed.

The Dimensional Analysis that we apply needs to make the following
distinctions. We need to know beforehand the set of fundamental quantities,
in this case it is solely the cosmic time $t$ as is detached of the
homogeneity and isotropy of the model and to distinguish the set of
constants, universal and unavoidable or characteristic, that in this case,
they are respectively the speed of the light $c$ and the constant of
integration $A_\omega $ that depending on the state equation that is
imposed, it will have different dimensions and physical meaning.

In a previous paper (\cite{T}) was calculated the dimensional base of this
type of models, being this $B=\left\{ L,M,T,\theta \right\} $ where $\theta $
represents the dimension of the temperature. The corresponding dimensions of
each quantity (with respect to this base $B$) are%
$$
\left[ t\right] =T\quad \left[ c\right] =LT^{-1}\quad \left[ A_\omega
\right] =L^{2+3\omega }MT^{-2} 
$$
All the quantities that we are going to calculate we will make it
exclusively in function of the cosmic time $t$ and of the unavoidable
constants, $c$ and $A_\omega $ with respect to the dimensional base $%
B=\left\{ L,M,T,\theta \right\} .$

\section{{\bf Solutions through D.A}.}

We are going to calculate through dimensional analysis D.A. the variation of 
$G(t)$ in function so much of $t$ as of the temperature $\theta ,$ $G(\theta
)$ (\cite{Z}), the energy density $\rho (t),$ the radius of the universe $%
f(t),$ the temperature $\theta (t)$, as well as to entropy $s(t)$ and the
entropy density $S(t)$ and finally $\Lambda (t).$

\subsection{\bf Calculation of G(t)}

As we have indicated above, we are going to accomplish the calculate
applying the Pi theorem. The quantities that we consider are $%
G=G(t,c,A_\omega )$ with respect to the dimensional base $B=\left\{
L,M,T,\theta \right\} .$ We know that $\left[ G\right] =L^3M^{-1}T^{-2}$
them:%
$$
\left( 
\begin{array}{rrrrr}
& G & t & c & A_\omega \\ 
L & 3 & 0 & 1 & 2+3\omega \\ 
M & -1 & 0 & 0 & 1 \\ 
T & -2 & 1 & -1 & -2 
\end{array}
\right) 
$$
\begin{equation}
\label{r1}G\propto \frac{t^{1+3\omega }c^{5+3\omega }}{A_\omega } 
\end{equation}
the cases more well-known are:

\begin{enumerate}
\item  Radiation predominance $\omega =\frac 13$%
\begin{equation}
\label{g1}G\propto \frac{t^2c^6}{A_\omega }\qquad \qquad G\propto t^2 
\end{equation}
In this expresion, if we want to calculate the value of the constant $%
A_\omega $ we can take into account the numerical values of the rest of the
quantities, that is to say: $G\approx 10^{-10.1757}m^3kg^{-1}s^{-2}$ , $%
t\approx 10^{20}s$ and $c\approx 10^{8.5}ms^{-1}$ with this values we obtain
that: \frame{$A_\omega \approx 10^{100.5}m^3kgs^{-2}$}

\item  Matter predominance $\omega =0$%
\begin{equation}
\label{g2}G\propto \frac{tc^5}{A_\omega }\qquad \qquad G\propto t
\end{equation}
\end{enumerate}

This result was obtained for Milne in 1935. If we consider $\rho $ as mass
density then $A_\omega $ represents the mass of the universe, the equation (%
\ref{r1}) is in this case to: 
\begin{equation}
\label{g3}G\propto \frac{tc^3}{A_\omega }\qquad \qquad G\propto t 
\end{equation}
this relation verify the kwon Sciama's formulate: $\rho Gt^2\approx 1$
(about the Mach's principle).

\begin{itemize}
\item  Furthermore if we take into account the numerical values of the
constant and the quantity $t$ we obtain the nowadays value of $G\approx
10^{-10.1757}m^3kg^{-1}s^{-2}$ i.e. $t\approx 10^{20}s,$ $c\approx
10^{8.5}ms^{-1}$ and $A_\omega \approx 10^{56}kg.$

\item  If $\omega =-\frac 13$ then $G=const.$ though this possibility is
physically unrealistic. If $\omega <-\frac 13$ then $G$ will vary in a way
inversely proportional to time.
\end{itemize}

If we want to relate $G$ with $\theta $ (ver \cite{Z}) the solution that the
D.A. provide is: $G=G(t,c,A_\omega ,a,\theta ).$ We need to introduce a new
constant, in this case thermodynamic, to relate the temperature to the rest
of quantities. The same result is obtained if we consider $k_B.$%
\begin{equation}
\label{r2}G\propto \frac{c^4\left( a\theta ^4\right) ^{\frac{\omega -1}{%
3(\omega +1)}}}{A_\omega ^{\frac 1{3(\omega +1)}}} 
\end{equation}
if $\omega =\frac 13$ 
$$
G\propto \frac{c^4}{\left( A_\omega a\right) ^{\frac 12}\theta ^2}\qquad
\qquad G\propto \theta ^{-2} 
$$
.

\subsection{\bf \ Calculation of energy density $\rho (t).$}

$\rho =\rho (t,c,A_\omega )$ that with respect to the base $B:\left[ \rho
\right] =L^{-1}MT^{-2}$%
\begin{equation}
\label{r3}\rho \propto \frac{A_\omega }{\left( ct\right) ^{3(\omega +1)}} 
\end{equation}

\begin{enumerate}
\item  If $\omega =1/3$ radiation predominance 
\begin{equation}
\label{d1}\rho \propto \frac{A_\omega }{\left( ct\right) ^4}\text{ \qquad }%
i.e.\text{ \quad }\rho \propto t^{-4}
\end{equation}
If we take into account the value of $A_\omega $ from equt. (\ref{g1}) we
can see easily that the value of the energy density correspond with the
nowadays accepted i.e. $\rho \approx 10^{-13.379}Jm^{-3}$

\item  If $\omega =0$ matter predominance 
\begin{equation}
\label{d2}\rho \propto \frac{A_\omega }{\left( ct\right) ^3}\text{ \qquad }%
i.e.\text{ \qquad }\rho \propto t^{-3}
\end{equation}
\end{enumerate}

\subsection{\bf Calculation of the radius of the Universe $f(t).$}

$f=f(t,c,A_\omega )$ where $\left[ f\right] =L\Longrightarrow $%
\begin{equation}
\label{r4}f\propto ct\qquad \qquad \qquad f\propto t 
\end{equation}
it does not depend on $A_\omega $ i.e. it does not depend on the state
equation, in both models, radiation and matter predominance, the radius of
the Universe is the same. We can observe that:%
$$
q=-\frac{f^{\prime \prime }f}{\left( f^{\prime }\right) ^2}=0 
$$
$$
H=\frac{f^{\prime }}f=\frac 1t 
$$
$$
d_H=ct\lim _{t_0\rightarrow 0}\int_{t_0}^t\frac{dt^{\prime }}{f(t^{\prime })}%
=\infty 
$$
Thus, the model has{\bf \ no horizon problem} because $d_H$ diverges for $%
t_0\rightarrow 0.$

\subsection{\bf Calculation of the temperature $\theta (t).$}

$\theta =\theta (t,c,A_{\omega ,}a)$ where the dimensional equation of the
temperature is $\left[ \theta \right] =\theta $ and $a$ is the radiation
constant. 
\begin{equation}
\label{r5}a^{\frac 14}\theta \propto \frac{A_\omega ^{\frac 14}}{\left(
ct\right) ^{\frac 34(1+\omega )}} 
\end{equation}
if $\omega =\frac 13\Rightarrow $%
\begin{equation}
\label{t1}a^{\frac 14}\theta \propto \frac{A_\omega ^{\frac 14}}{\left(
ct\right) }\text{ \qquad \qquad }\theta \propto t^{-1} 
\end{equation}
If we take into account the value of $A_\omega $ from equt. (\ref{g1}) we
can see easily that the value of the temperature correspond with the
nowadays accepted i.e. $\theta \approx 10^{0.4361}K$ or $\theta \approx
2.73^{\circ }K$ where the radiation constant take the value $a\approx
10^{-15.1211}Jm^{-3}K^{-4}.$ We can show also that with this behaviour we
obtain the well-known result:%
$$
\rho =a\theta ^4 
$$

\subsection{\bf Calculation of the Entropy}

$s=s(c,A_{\omega ,}a)$ here we do not consider the time since upon imposing $%
div(T_{ij})=0$ the equations that we are considering are adiabatics i.e.
they do not depend on the time. $\left[ s\right] =L^2MT^{-2}\theta ^{-1}$ 
\begin{equation}
\label{r6}s\propto \left( A_\omega ^3ac^{3(1-3\omega )}\right) ^{\frac 14} 
\end{equation}
if $\omega =\frac 13\Rightarrow $%
\begin{equation}
\label{m1}s\propto \left( A_\omega ^3a\right) ^{\frac 14}\text{ \qquad
\qquad }s=const. 
\end{equation}

\subsection{{\bf Entropy Density }$S(t).$}

$S=S(t,c,A_{\omega ,}a)$ \qquad $\left[ S\right] =L^{-1}MT^{-2}\theta ^{-1}$ 
\begin{equation}
\label{r7}S\propto \frac{\left( A_\omega ^3a\right) ^{\frac 14}}{\left(
ct\right) ^{\frac 94(1+\omega )}} 
\end{equation}
if $\omega =\frac 13\Rightarrow $%
\begin{equation}
\label{m2}S\propto \frac{\left( A_\omega ^3a\right) ^{\frac 14}}{\left(
ct\right) ^3}\qquad \qquad S\propto t^{-3} 
\end{equation}

\subsection{\bf Calculation of the cosmological constant $\Lambda $$(t)$}

$\Lambda =\Lambda (t,c,A_\omega );$ $\left[ \Lambda \right] =L^{-2}$%
\begin{equation}
\label{r8}\Lambda \propto \frac 1{c^2t^2} 
\end{equation}
it does not depend on $A_\omega $ i.e. it does not depend on the state
equation, for this reason this quantity show the same behavior in both
models. We can show that if $t\approx 10^{20}s$ and $c\approx
10^{8.4}ms^{-1} $ we obtain the at present admited value of the so called
cosmological constant i.e. $\Lambda \approx 10^{-56}m.$

\section{\bf Summary and conclusions.}

We have solved through dimensional analysis a FRW model with $k=0$ and with
conservation of the energy-momentum tensor where $G$ and $\Lambda $ are
variables with respect to the time $t$. We have found that $G\propto
t^{1+3\omega }$ i.e. depend on the equation of state. If $\omega =\frac 13$
we obtain that $G\propto t^2$ and if $\omega =0$ , then, $G\propto t$ while, 
$\Lambda \propto t^{-2}$ independently from the state equation that we
impose. Only it appears $G=const.$ when $\omega =-\frac 13$ state equation
this physically unrealistic. The radius of the universe either depends on
the state equation and it is directly proportional to the time $f\propto t$
,with this radius, the model does not has the problem of the horizon and
appears with $q=0$ i.e. an always expansive universe. This result are not in
disagreement with the values nowadays accepted.

\subparagraph{\bf ACKNOWLEDGEMENTS.}

I wish to thank Prof. M. Casta\~ns for suggestions and enlightening
discussions

\end{document}